\documentclass[11pt,letterpaper]{article}
\usepackage{mathtools}
\usepackage{axodraw2}
\usepackage{tikz}
\usepackage{tikz-feynman}
\usepackage{amscd}
\usepackage{amsmath}
\usepackage{slashed}
\usepackage{gensymb}
\usepackage{amssymb}
\usepackage{bm}
\usepackage{jheppub} 
\usepackage[normalem]{ulem} 
\allowdisplaybreaks 

\newcommand{\bsll}{$b \rightarrow s \ell^+ \ell^-$}

\newcommand{\smelli}{{\tt smelli2.4.1}}
\newcommand{\flavio}{{\tt flavio2.6.1}}
\newcommand{\wilson}{{\tt wilson2.3.2}}

\newcommand{\K}{\mathcal{K}}

\newcommand{\improvement}{7.0}
\newcommand{\bestchisq}{38.2}

\title{\centering
Malaphoric {\boldmath $Z^\prime$} models for {\boldmath \bsll}\ anomalies}

\author{Ben Allanach}
\emailAdd{ben.allanach.work@gmail.com}
\author{and Nico Gubernari}
\emailAdd{nico.gubernari@gmail.com}
\affiliation{DAMTP, University of Cambridge, Wilberforce Road, Cambridge,
  CB3 0WA, United Kingdom}

\abstract{
  We study some phenomenological effects of kinetic mixing between the hypercharge field and a new $U(1)$ gauge field in specific $Z^\prime$ models that ameliorate the tensions between measurements and Standard Model predictions in $b \rightarrow s\ell^+ \ell^-$ decays.
  To this end, we rederive the dimension-6 SMEFT coefficients resulting from
  integrating out the kinetically-mixed (`malaphoric') $Z^\prime$
  field.
  The kinetic mixing provides a family-universal component to the couplings of the $Z^\prime$ field, which can improve fits to lepton flavour universality observables.
  We show how kinetic mixing improves
  the best fit of the $B_3-L_2$ model to $b\rightarrow s$ data by
  \improvement\ units of $\chi^2$ while remaining compatible with other
  relevant data sets such as electroweak precision observables and
  measurements of $e^+ e^- \rightarrow \ell^+ \ell^-$ at LEP2.
  }


\begin{document}

\maketitle
\flushbottom

\section{Introduction \label{sec:intro}}

The $B-$meson decays mediated by a $b \rightarrow s \ell^+\ell^-$ transition are a powerful tool to test the  validity  of the Standard Model (SM).
In the last decades, the flavour physics community has invested substantial resources to improve the precision of both SM predictions and experimental measurements of these decays to obtain stronger constraints on physics beyond the SM (BSM).
Whilst many such measurements are compatible with the SM~\cite{Gubernari:2022hxn,Parrott:2022zte}, those that are incompatible include the angular distribution and the branching ratio ($BR$)
in $B \rightarrow K^\ast \mu^+ \mu^-$
decays~\cite{LHCb:2020lmf,ATLAS:2018gqc,CMS:2017rzx,LHCb:2016ykl}, $BR(B \rightarrow K
\mu^+ \mu^-)$~\cite{LHCb:2014cxe,CMS:2024syx}, and $BR(B_s \rightarrow \phi \mu^+\mu^-)$~\cite{LHCb:2021zwz}.
These discrepant observables suffer from
relatively large theoretical errors due to hard-to-calculate non-local QCD effects. Nevertheless, several estimates suggest that these effects are too small to resolve the
discrepancies~\cite{Mutke:2024tww,Ball:2006eu,Isidori:2024lng,Gubernari:2020eft}. Two observables whose current measurements broadly
agree~\cite{LHCb:2022qnv} with SM predictions are $R_K$ and $R_{K^\ast}$, where
\begin{equation}
  R_{M}(q^2_\text{min},\ q^2_\text{max}) :=
  \frac{
    \int_{q^2_\text{min}}^{q^2_\text{max}} dq^2
    \frac{
      dBR(B \rightarrow M\mu^+\mu^-)}
      {dq^2}
  }{
    \int_{q^2_\text{min}}^{q^2_\text{max}} dq^2
    \frac{
      dBR(B \rightarrow M e^+e^-)}
      {dq^2}
  }, \label{Rdef}
\end{equation}
for a meson $M$.
This is important because, if one hypothesises BSM contributions to
affect the aforementioned discrepant decays in observables involving
di-muon pairs, the agreement of $R_K$ and $R_{K^\ast}$ with lepton flavour
universality (LFU) may also imply some
BSM effects in $b \rightarrow s e^+ e^-$ (plus $CP-$conjugate) decays.

We display a SM fit provided by\footnote{
  In \smelli\ and \flavio, the Wilson coefficients are run, converted and matched by
  \wilson~\cite{Aebischer:2018bkb}.
  We had \smelli\ first recalculate the
  covariances in the theoretical uncertainties, a step which has now been done
  for the user in {\tt smelli2.4.2}.}
\smelli~\cite{Aebischer:2018iyb} and \flavio~\cite{Straub:2018kue} to various
categories of observables
in Table~\ref{tab:SM}.
The categories are pre-defined in
\flavio\ (we have omitted certain categories which are not relevant for our analysis) and are:
\begin{itemize}
\item
  `quarks': this category includes many of the aforementioned discrepant observables in
  angular distributions of $B \rightarrow K^\ast \mu^+ \mu^-$ decays,  as well
  as branching ratios of $B \rightarrow K^{(\ast)} \mu^+ \mu^-$ in various
  bins. It also contains other observables which  agree with
  their SM prediction such as $\Delta M_s$, a measure of
  the amount of $B_s-\bar{B}_s$ mixing.
  $\Delta M_s$ constrains $Z^\prime$
  models that couple to $\bar b s$ and $b \bar s$ such as the
  model which we shall study.
\item
  `LEP2': this category constitutes differential cross-sections for $e^+ e^-$ collisions
  producing di-lepton pairs at the CERN LEP and LEP2 colliders. The
  calculation of the likelihood of these observables is described in
  Ref.~\cite{Allanach:2023uxz} where it was made available in a form that was
  interfaced with {\tt flavio} and {\tt smelli}. We will use identical
  computer code for this incorporation in the present  paper.
  Note that this is the only category not included in the original {\tt flavio} code.
\item
  `EWPOs': this category consists of the electroweak precision observables \emph{without} a
  family universality assumption. Thus, for example, $BR(Z \rightarrow \mu^+
  \mu^-)$ is measured separately to $BR(Z \rightarrow e^+e^-)$, even though
  the SM predicts them to be equal. This is appropriate for models which distinguish between different families, such as the ones we shall study.
\item
  `LFU': this contains various lepton flavour universality variables. Of
  particular interest are $R_K$ and $R_{K^\ast}$ from (\ref{Rdef})
  because they used~\cite{LHCb:2021trn} to exhibit
    significant deviations with respect to SM predictions (according to
    analyses by the LHCb experiment), but do so no longer~\cite{LHCb:2022qnv}.
\end{itemize}
\begin{table}
  \begin{center}
    \begin{tabular} {|c|cccc|c|} \hline
      set & quarks & EWPOs & LEP2 & LFU & global \\ \hline
      $\chi^2$ & 419. & 36.8 & 150. & 18.5 & 624. \\
      $N$      & 306 & 31    & 148  & 24  & 509 \\
      $p$      & $1.7 \times 10^{-5}$ & $.22$ & $.44$ & $.78$ & $3.4 \times 10^{-4}$ \\
    \hline
  \end{tabular}
  \end{center}
  \caption{SM fit to \flavio\ data sets (shown as the middle four column
    headings) used in the present paper. $\chi^2$ is the common
    chi-squared statistic, $N$ is the number of observables in the set and $p$
    is the
    $p-$value (here, of the SM hypothesis with the \flavio\ estimate of
    observables and uncertainties).
    \label{tab:SM}}
\end{table}
Table~\ref{tab:SM} shows that the SM quality-of-fit is poor. While the EWPOs, LEP2 and LFU measurements
are all compatible with the SM, the observables in the quarks category as
a whole are
distinctively incompatible with it. Many observables within the quarks category
\emph{are} compatible with SM predictions: the $p-$value omitting these is of
course lower than the one quoted, but we stick to the pre-defined categories in
\flavio\ in
order to evade accusations of a-posteriori bias.
Many of the observables in the quarks category have large theoretical
uncertainties in their predictions: these are taken into account with a
generous error budget in {\tt flavio}.
We shall next describe some recent attempts to ameliorate the fit to the observables
in the `quarks' category, while simultaneously still fitting the other
categories of observables reasonably well.

In Refs.~\cite{Bonilla:2017lsq,Alonso:2017uky,Allanach:2020kss}, the
hypothesis of a TeV-scale $Z^\prime$ with certain
family-dependent couplings was investigated to explain
$B-$decay data.
Such a $Z^\prime$ results from a hypothesised $U(1)_X$
gauge symmetry with family-dependent charges for the SM fermions
(plus three
right-handed neutrinos) which is
spontaneously broken at the TeV scale.
The fermion charge assignment here\footnote{Other family dependent
$U(1)$ symmetries were also proposed for this purpose, for example involving
$L_2-L_3$~\cite{Altmannshofer:2014cfa,Crivellin:2015lwa} or
third family hypercharge~\cite{Allanach:2018lvl}.} was
$B_3-L_2$, third-family baryon number minus second-family lepton
number. Mixing between left-handed bottom and strange quark fields
provides a coupling between the $Z^\prime$ field and $\bar b s+h.c.$, whereas
the
$Z^\prime$ field couples directly to di-muon pairs because the muon
field's $U(1)_X$ charge is non-zero.
However, as described above,
$R_K$ and $R_{K^\ast}$ became
compatible with their SM predictions more recently than the earlier
$B_3-L_2$ papers
(i.e.\ Refs.~\cite{Bonilla:2017lsq,Alonso:2017uky,Allanach:2020kss}).
Given the absence of a direct coupling between the $Z^\prime$ and di-electron pairs in the $B_3-L_2$ model, the parameter space region that successfully explains the discrepant muon observables no longer adequately describes the updated experimental values of $R_K$ and $R_{K^\star}$~\cite{Allanach:2022iod}.
While this region still represents a significant improvement over the SM, it is no longer able to reconcile both sets of measurements simultaneously.

To the best of our knowledge, $Z^\prime$ explanations of the
\bsll\ anomalies, including the $B_3-L_2$ model,
have all neglected kinetic mixing
between the $U(1)_X$
gauge boson field and the hypercharge gauge boson field.
Such a mixing
term in the Lagrangian density is gauge invariant and thus allowed. Even
if it is not present, it will be generically induced by loop
effects. However, if one has a reason to suppose that the mixing effects
are zero at a certain larger scale (for example because $SU(3)\times SU(2)
\times U(1)_Y \times U(1)_X$ becomes embedded in some appropriate semi-simple group at that
scale), then loop corrections will be small and neglecting the kinetic
mixing term should be a good approximation.

In the present paper we shall, for the first time,
explore the possibility that the kinetic mixing term is
sizeable, dubbing the resulting models `malaphoric $Z^\prime$ models'.
The kinetic mixing will introduce a family universal component to the couplings of the $Z^\prime$ field to leptons (as well as to other fermionic fields), and this can improve the fit to $R_K$ and $R_{K^\ast}$ whilst still fitting the other discrepant flavour data\footnote{Mixing with the hypercharge boson will affect the EWPOs, and the $Z^\prime$-di-electron coupling will affect the LEP2 data.
Thus, one can understand why the four categories of observables in
Table~\ref{tab:SM} are important for our analysis.}.
We shall take the example of the
$B_3-L_2$ model and add the effects of sizeable kinetic mixing in order to
examine quantitatively how the fit to measurements changes.

In order to calculate such a change, we must inform
\smelli\ and \flavio{}
of the BSM effects
in the form of SM effective field theory (SMEFT)
dimension-6
coefficients, resulting from integrating out the $Z^\prime$ from the effective
field theory. We shall present this calculation in \S\ref{sec:coeffs}, for
generic fermionic charges.
The fits to the $B_3-L_2$ model, which we review in \S\ref{sec:model}, will be presented in \S\ref{sec:fits}, both including the effects of kinetic mixing and neglecting it, for the purposes of illustration.
We shall summarise and conclude in \S\ref{sec:conc}.

\section{SMEFT Coefficients \label{sec:coeffs}}

The SMEFT coefficients of generic $Z^\prime$ models including the effects of
kinetic mixing were recently calculated for the first time in
Ref.~\cite{Dawson:2024ozw}. We shall re-derive the dimension-6 coefficients,
providing an independent check.
As in Ref.~\cite{Dawson:2024ozw}, we start with the part of the Lagrangian that contains the additional $U(1)_X$ gauge field\footnote{
  In our discussions, we shall discriminate between the $X^\mu$
  field, which is the (potentially) kinetically mixed gauge eigenstate spin-1 vector boson,
  versus the propagating eigenstate $Z^\prime$.
}$X^\mu$ and the hypercharge gauge boson field $B^\mu$:
\begin{equation}
  {\mathcal L}_{XB} = -\frac{1}{4} X_{\mu\nu} X^{\mu \nu} +
  g_X^2 X_H^2 |H^\dag H|  X_\mu X^\mu +
  \frac{1}{2} M_X^2
  X_\mu   X^\mu
  - \frac{\epsilon}{2} B_{\mu \nu}X^{\mu \nu}- X_\mu J^\mu - B_\mu j^\mu,
  \label{eq:LXB}
\end{equation}
where $g_X$ is the $U(1)_X$ gauge coupling and $X_H$ is the charge of the
Higgs doublet under $U(1)_X$.
Here, the $U(1)$ field strengths are defined as $X_{\mu\nu}:=\partial_\mu X_\nu - \partial_\nu X_\nu$ and $B_{\mu \nu}:=\partial_\mu
 B_\nu - \partial_\nu B_\mu$.
 The term proportional to
 $\epsilon$ in (\ref{eq:LXB}) is the one responsible for the kinetic mixing
 and $|\epsilon|\leq 1$ in order for the kinetic terms of both
 neutral gauge boson eigenstates to have the correct sign~\cite{Babu:1997st,Cheng:2024hvq}.
We have defined
  \begin{equation}
J_\mu = i g_X X_H\left[ H^\dagger D_\mu H - (D_\mu H)^\dag H\right]+g_X \sum_{\psi^\prime}  X_{\psi^\prime} \overline{\psi^\prime}
\gamma_\mu \psi^\prime, \label{eq:Xcurr}
\end{equation}
where $X_{\psi^\prime}$ is the charge under $U(1)_X$ of chiral fermionic weak
eigenstate $\psi^\prime$.
The sum in (\ref{eq:Xcurr}) and in the equation below runs over all chiral fermionic fields $\psi^\prime$ in the SM gauge group, i.e. $q_i^\prime$, $l_i^\prime$, $e_i^\prime$, $d_i^\prime$ and $u_i^\prime$.
The hypercharge current is
\begin{equation}
j_\mu = i g^\prime Y_H \left[ H^\dag D_\mu H - (D_\mu H)^\dag H \right]+ g^\prime \sum_{\psi^\prime}  Y_\psi \overline{\psi^\prime} \gamma_\mu \psi^\prime.
\label{eq:hypCurr}
\end{equation}
Here $Y_H$ is the hypercharge of the Higgs field, $Y_\psi$ is the hypercharge of the fermionic field $\psi^\prime$ and $g^\prime$ is the hypercharge gauge coupling.
The electroweak covariant derivative is defined as
\begin{equation}
D_\mu = \partial_\mu + i g^\prime Y_H B_\mu + i g W_\mu^a T^a,
\end{equation}
where $g$ is the SM $SU(2)$ gauge coupling, $W^a_\mu$ are three $SU(2)$
gauge fields and $T^a$ are the three generators of $SU(2)$ in
the fundamental representation.
Our conventions for $U(1)$ charges along with the other quantum numbers of the fields are tabulated in Table~\ref{tab:charges}.
\begin{table}
  \begin{center}
  \begin{tabular} {|c|cccccc|} \hline
    Field $\phi$ & $q_i^\prime$ & $l_i^\prime$ & $e_i^\prime$ & $d_i^\prime$ & $u_i^\prime$ & $H$\\ \hline
    $SU(3)$      & 3 & 1 & 1 & 3 & 3 & 1 \\
    $SU(2)$      & 2 & 2 & 1 & 1 & 1 & 2 \\
    $Y_\phi$     & 1/6   & -1/2  & -1    & -1/3  & 2/3   & 1/2 \\
    $X_\phi$ & $X_{q_i}$ & $X_{l_i}$ & $X_{e_i}$ &  $X_{d_i}$ & $X_{u_i}$ & $X_H$ \\
    \hline
  \end{tabular}
  \end{center}
  \caption{Representation of gauge eigenstates of matter and Higgs fields under $SU(3)\times SU(2)
  \times U(1)_Y \times U(1)_X$.  The
  right-handed neutrino fields are assumed to be super-heavy and thus to play
  a negligible role in the collider phenomenology investigated
  here. \label{tab:charges}}
\end{table}

In order to calculate SMEFT coefficients, we integrate out the heavy
field $X_\mu$ yielding (up to yet higher dimension operators) the
dimension-6 Lagrangian density terms
\begin{equation}
\mathcal{L}_6 = -\frac{1}{2 M_{X}^2} {J}_\mu {J}^\mu -
\frac{\epsilon}{M_{X}^2} (\partial_\nu B^{\mu \nu}) {J}_\mu -
\frac{\epsilon^2}{2 M_{X}^2} (\partial_\nu B^{\mu \nu}) (\partial^\rho
  B_{\mu \rho}). \label{eq:L6}
\end{equation}
The procedure to integrate out the $X_\mu$ field is standard and is reported in  Appendix~\ref{app:intout}.
Applying the 4-dimensional equation of motion (again, equivalently
neglecting yet higher dimension operators)
$j^\mu = -\partial_\nu B^{\mu \nu}$
to (\ref{eq:L6}), we obtain the dimension-6 terms
\begin{equation}
\mathcal{L}_6 = -\frac{1}{2 M_{X}^2} \left(J_\mu - \epsilon j_\mu \right)
\left(J^\mu - \epsilon j^\mu\right), \label{eq:Jj}
\end{equation}
which implicitly encodes the various SMEFT operators and coefficients.
In order to extract the SMEFT coefficients in the down-aligned Warsaw basis~\cite{Grzadkowski:2010es}
we expand (\ref{eq:Jj}) into the unprimed fermion mass eigenstates, for
example
\begin{equation}
  q_{i}^\prime = (V_{d_L})_{ij} q_{j}.
\end{equation}
In the \emph{down-aligned} Warsaw basis, the $q_{i}^\prime$ fields are
assumed to
have already been rotated in
family space: $q_{i}^\prime = (V_{q})_{ij} \tilde q_{j}$ from an initial
field basis $\tilde q_{j}$ such that $q_{i}^\prime=(V_{ik}
u_{L_k}^\prime,\ d_{L_i}^\prime)$, where $(V_{q})_{ij}$ are elements of some unitary 3 by
3 matrix and $V$ is the CKM matrix~\cite{ParticleDataGroup:2022pth}.
The mass eigenstates of
the $u_i$, $d_i$, $e_i$ fields are assumed to
coincide with those of the concomitant gauge eigenstates, i.e.\ $u_i=u_i^\prime$, $d_i=d_i^\prime$, $e_i=e_i^\prime$.
$e_{L_i}$ is also assumed to coincide with $e_{L_i}^\prime$ whereas $V_{d_L}$
necessarily will contain some family mixing (in order to facilitate certain
$b\rightarrow s$ transitions), to be defined in more detail below.

Expanding (\ref{eq:Jj}) in terms of the fields
leads to the calculation of the SMEFT Wilson coefficients in the down-aligned
Warsaw basis~\cite{Grzadkowski:2010es}.
The dimension-6 SMEFT coefficients we present are initially in a
\emph{redundant} basis, where the family indices $i$, $j$, $k$ and $l$ take values $\in \{1,2,3\}$ and
for example both $C_{ll}^{iijj}$ and
$C_{ll}^{jjii}$ are non-zero\footnote{From now on, repeated indices are
left unsummed unless the summation is explicit.}, even though they must be
equal. In our
numerics, these are carefully converted to the \emph{non}-redundant basis
defined in Ref.~\cite{Celis:2017hod} used in the Wilson Coefficient Exchange
Format ({\tt WCXF})~\cite{Aebischer:2017ugx}. We shall explain (following each set of
coefficients) how the expressions change in conversion to the {\tt WCXF} format,
if at all. We shall list only the \emph{non-zero} dimension-6 SMEFT coefficients,
starting with the 4-fermion operators in their mass eigenbasis
\begin{align}
  C_{ll}^{iijj} &= -\frac{1}{2 M_{X}^2} \left(g_X X_{l_i} - g^\prime
    \epsilon Y_l \right) \left( g_X X_{l_j} - g^\prime
    \epsilon Y_l \right), \label{Cll} \\
  C_{ee}^{iijj} &= -\frac{1}{2 M_{X}^2} \left(g_X X_{e_i} - g^\prime
    \epsilon Y_e \right) \left( g_X X_{e_j} - g^\prime
    \epsilon Y_e \right), \\
  C_{uu}^{iijj} &= -\frac{1}{2 M_{X}^2} \left(g_X X_{u_i} - g^\prime
    \epsilon Y_u \right) \left( g_X X_{u_j} - g^\prime
    \epsilon Y_u \right), \\
  C_{dd}^{iijj} &= -\frac{1}{2 M_{X}^2} \left(g_X X_{d_i} - g^\prime
    \epsilon Y_d \right) \left( g_X X_{d_j} - g^\prime
    \epsilon Y_d \right). \label{Cdd}
\end{align}
To convert to the {\tt WCXF} non-redundant basis, one requires $j\ge i$ and
multiplies each of the right-hand sides of (\ref{Cll})-(\ref{Cdd}) by a factor
$(2-\delta_{ij})$.
In order to write down $C_{qq}^{(1)}$ in a succinct form, we first define
\begin{equation}
  \Xi_{ij} := - Y_q
  g^\prime \epsilon \delta_{ij} +
  g_X \sum_k (V_{d_L}^\dag)_{ik} X_{q_k} (V_{d_L})_{kj}
  \,,
\end{equation}
so that
\begin{equation}
  (C_{qq}^{(1)})^{ijkl} = -\frac{1}{2 M_{X}^2} \Xi_{ij} \Xi_{kl}. \label{Cqq}
\end{equation}
Here, for this one set of SMEFT Wilson coefficients, converting to {\tt WCXF}
format is slightly more involved. We cycle
through all possible values of $\{i,j,k,l\}\in \{1,2,3\}$, adding (\ref{Cqq})
(if it is
equivalent) to the relevant coefficient of the
27 non-redundant $(C_{qq}^{(1)})^{ijkl}$ defined in
Ref.~\cite{Aebischer:2017ugx}\footnote{For
details, see the main program {\tt kinetic\_mixing.py} provided in the
ancillary directory of the {\tt arxiv} version of this paper.}. This automatically
reproduces symmetry factors from the summands.
We also have the non-identical 4-fermion operators (leaving  the prime in
$X_{\psi^\prime}$ implicit for brevity's sake)
\begin{align}
  C_{le}^{iijj} &= -\frac{1}{M_{X}^2} \left(g_X X_{l_i} - g^\prime
    \epsilon Y_l \right) \left( g_X X_{e_j} - g^\prime
    \epsilon Y_e \right), \label{Cle} \\
  C_{lu}^{iijj} &= -\frac{1}{M_{X}^2} \left(g_X X_{l_i} - g^\prime
    \epsilon Y_l \right) \left( g_X X_{u_j} - g^\prime
    \epsilon Y_u \right), \label{Clu} \\
  C_{ld}^{iijj} &= -\frac{1}{M_{X}^2} \left(g_X X_{l_i} - g^\prime
    \epsilon Y_l \right) \left( g_X X_{d_j} - g^\prime
    \epsilon Y_d \right), \label{Cld} \\
  C_{eu}^{iijj} &= -\frac{1}{M_{X}^2} \left(g_X X_{e_i} - g^\prime
    \epsilon Y_e \right) \left( g_X X_{u_j} - g^\prime
    \epsilon Y_u \right), \label{Ceu} \\
  C_{ed}^{iijj} &= -\frac{1}{M_{X}^2} \left(g_X X_{e_i} - g^\prime
    \epsilon Y_e \right) \left( g_X X_{d_j} - g^\prime
    \epsilon Y_d \right), \label{Ced} \\
  (C_{ud}^{(1)})^{iijj} &= -\frac{1}{M_{X}^2} \left(g_X X_{u_i} - g^\prime
    \epsilon Y_u \right) \left( g_X X_{d_j} - g^\prime
    \epsilon Y_d \right), \label{Cud1}
  \end{align}
which are already in the {\tt WCXF} format.
We also have the coefficients of 4-fermion operators involving only a single bilinear of the
$q_i$ fields
\begin{align}
  (C_{lq}^{(1)})^{kkij} &= -\frac{1}{M_{X}^2} \left(g_X X_{l_k} - g^\prime
  \epsilon Y_l \right) \Xi_{ij}, \label{Clq1} \\
  C_{qe}^{ijkk} &= -\frac{1}{M_{X}^2}  \Xi_{ij} \left(g_X X_{e_k} - g^\prime
  \epsilon Y_e \right), \label{Cqe} \\
  (C_{qu}^{(1)})^{ijkk} &= -\frac{1}{M_{X}^2}  \Xi_{ij}\left(g_X X_{u_k} - g^\prime  \epsilon Y_u \right) , \label{Cqu1} \\
  (C_{qd}^{(1)})^{ijkk} &= -\frac{1}{M_{X}^2}  \Xi_{ij}\left(g_X X_{d_k} - g^\prime  \epsilon Y_d \right), \label{Cqd1}
\end{align}
where in the {\tt WCXF}, (\ref{Clq1})-(\ref{Cqd1}) should have $j \ge i$ and
then the Hermitian conjugate terms with $j<i$ are
implied.
This condition also applies to
\begin{equation}
  (C_{\varphi q}^{(1)})_{ij} = -\frac{1}{M_{X}^2} (g_XX_H-g^\prime
  \epsilon Y_H)
  \Xi_{ij}. \label{Cphiq_1}
\end{equation}
No conversion is necessary to obtain the remaining non-zero coefficients in the {\tt
  WCXF}:
\begin{align}
 (C_{\varphi l}^{(1)})^{ii} &= -\frac{1}{M_X^2} (g_X X_H - g^\prime \epsilon Y_H) \left(g_X X_{l_i} - g^\prime  \epsilon Y_l \right), \label{Cphil1} \\
  C_{\varphi e}^{ii} &= -\frac{1}{M_X^2} (g_X X_H - g^\prime \epsilon Y_H) \left(g_X X_{e_i} - g^\prime  \epsilon Y_e \right), \label{Cphie} \\
  C_{\varphi d}^{ii} &= -\frac{1}{M_X^2} (g_X X_H - g^\prime \epsilon Y_H) \left(g_X X_{d_i} - g^\prime  \epsilon Y_d \right), \label{Cphid} \\
  C_{\varphi u}^{ii} &= -\frac{1}{M_X^2} (g_X X_H - g^\prime \epsilon Y_H) \left(g_X X_{u_i} - g^\prime  \epsilon Y_u \right), \label{Cphiu} \\
  C_{\varphi \Box} &= - \frac{1}{2 M_X^2} \left( {g_X X_H-\epsilon g^\prime}Y_H\right)^2, \label{Cboxphi} \\
  C_{\varphi D} &= - \frac{2}{M_X^2}\left(g_X X_H - \epsilon {g^\prime}Y_H\right)^2. \label{CphiD}
\end{align}
This completes the list of dimension-6 non-zero SMEFT Wilson coefficients resulting from
integrating out the $X^\mu$ field.
We shall now introduce a particular model for studying the effects of kinetic mixing.

\section{$B_3-L_2$ Model \label{sec:model}}
We now briefly introduce the $B_3-L_2$
model~\cite{Bonilla:2017lsq,Alonso:2017uky,Allanach:2020kss}. This model was
devised so that the effects of the TeV-scale $Z^\prime$ field could explain
the \bsll\ anomalies. The $X:=B_3-L_2$ $U(1)_X$ charge assignments are displayed in
Table~\ref{tab:b3l2}. We have set $X_H=0$ because of the need to be
able to write
down a top Yukawa coupling in the model, given the other $X$ charges. Such a Yukawa coupling, being of
order one, should be allowed in the set of renormalisable dimension-4
Lagrangian terms.
\begin{table}
  \begin{center}
    \begin{tabular} {|c|cccccccc|} \hline
      Field $\phi$ & $q_i^\prime$ & $l_i^\prime$ & $e_i^\prime$ & $d_i^\prime$
      & $u_i^\prime$ & $\nu_i$ & $H$ & $\theta$\\ \hline
      $X_\phi$ & $\delta_{i3}$ & $-3\delta_{i2}$ & $-3\delta_{i2}$ &  $\delta_{i3}$ &
      $\delta_{i3}$ & $-3\delta_{i2}$ & $0$ & $+1$ \\
      \hline
    \end{tabular}
  \end{center}
  \caption{\label{tab:b3l2} $X$ charge assignments of fields in the
    $B_3-L_2$ model. The flavon $\theta$ is a SM-singlet complex scalar
    field which is assumed to acquire a TeV-scale vacuum expectation value.}
\end{table}
The $U(1)_X$ symmetry is spontaneously broken by the flavon $\theta$, which is
assumed to acquire a TeV-scale vacuum expectation value $\langle \theta \rangle$, resulting in a
massive spin-1 bosonic field $X^\mu$ with Lagrangian mass parameter
\begin{equation}
  M_X = g_X \langle \theta \rangle.
\end{equation}
The model explains why $|V_{td}|$, $|V_{ts}|$, $|V_{ub}|$ and $|V_{cb}|$ are
all small compared to unity: they are zero in the unbroken $U(1)_X$ limit but
the limit receives small corrections from the spontaneous breaking of the
gauge symmetry. In a similar vein, the model postdicts
a suppression in flavour changing effects between the first two families of
charged lepton. Searches for $\mu \rightarrow e \gamma$ confirm that the
process is much suppressed, since an upper bound on the branching ratio of
smaller than $10^{-12}$ has been placed~\cite{ParticleDataGroup:2024cfk}.

The dimension-6 SMEFT coefficients resulting from integrating the $Z^\prime$
of the $B_3-L_2$ model were calculated in Ref.~\cite{Allanach:2022iod}
in the $\epsilon=0$ limit.
We have checked that these coefficients agree  in that limit with
the expressions presented in \S\ref{sec:coeffs}.

\begin{figure}
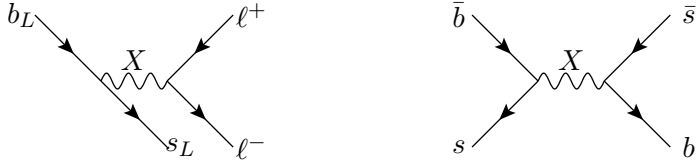

  \begin{center}
    \begin{axopicture}(80,55)(-5,-5)
      \Line[arrow](0,50)(25,25)
      \Line[arrow](25,25)(50,0)
      \Line[arrow](75,50)(50,25)
      \Line[arrow](50,25)(75,0)
      \Photon(25,25)(50,25){3}{3}
      \Text(37.5,33)[c]{$X$}
      \Text(55,0)[c]{${s_L}$}
      \Text(-5,50)[c]{$b_L$}
      \Text(82,50)[c]{$\ell^+$}
      \Text(82,0)[c]{$\ell^-$}
    \end{axopicture}
    \hskip 3cm
    \begin{axopicture}(80,55)(-5,-5)
      \Line[arrow](0,50)(25,25)
      \Line[arrow](25,25)(0,0)
      \Line[arrow](75,50)(50,25)
      \Line[arrow](50,25)(75,0)
      \Photon(25,25)(50,25){3}{3}
      \Text(37.5,33)[c]{$X$}
      \Text(-5,0)[c]{${s}$}
      \Text(-5,50)[c]{$\bar b$}
      \Text(82,50)[c]{$\bar s$}
      \Text(82,0)[c]{$b$}
    \end{axopicture}  \end{center}
  \caption{\label{fig:NCBAs} Feynman diagrams of the leading BSM
    contribution in the $B_3-L_2$ model to: \bsll\ observables (left),
    $B_s-\bar{B}_s$ mixing (right).
  }
\end{figure}

The left-handed down quark mixing matrix is assumed to have the form
\begin{equation}
  V_{d_L} = \begin{pmatrix}
    1 & 0 & 0 \\
    0 & \cos \theta_{sb} & \sin \theta_{sb} \\
    0 & -\sin \theta_{sb} & \cos \theta_{sb} \\
    \end{pmatrix}. \label{VdL}
  \end{equation}
  In order to predict non-SM $b\rightarrow s$ mixing effects and explain some
  of the \bsll\ anomalies as depicted in the left-hand panel of
  Fig.~\ref{fig:NCBAs}, one requires
  that $\theta_{sb} \neq 0$.
  This adds a tree-level flavour changing neutral
  current mediated by $X_\mu$. $U(1)_X$ explanations of \bsll\ anomalies such
  as the $B_3-L_2$ model
  induce a BSM contribution to $B_s-\bar{B}_s$ mixing, as
  shown in the right-hand panel of the same figure.
$B_s-\bar{B}_s$ mixing effects broadly agree with SM predictions and it
is important that the measurable observable relevant to this ($\Delta M_s$) is
included in any flavour fits or constraints upon the model.
\flavio\ computes this observable and it is included in our fits in the
`quarks' category.

We thus have strong assumptions for the mixing matrices such as $V_{d_L}$ but we have not explicitly done the specific ultra-violet model
  building to postdict them. We intend that this ansatz is to be understood to be
  approximate only (for example the zeroes in (\ref{VdL}) could contain
  numbers whose magnitude is much smaller than unity), and could come
  about from more detailed fermion mass and
  mixing modelling. For an indication of how
  such an ansatz might be achieved using the
  Froggatt-Nielsen mechanism, see
  Appendix A of Ref.~\cite{Allanach:2019iiy}.
  Leaving such detailed model-building considerations aside,
we now go on to describe
these fits and present the result of them.

\section{Fits \label{sec:fits}}
The SMEFT coefficients are functions of the four BSM
parameters:
$M_X$, $\epsilon$, $g_X$ and $\theta_{sb}$. However,
notice that the SMEFT coefficients (\ref{Cll})-(\ref{CphiD}) only depend
upon three free input BSM-parameter \emph{combinations}, which can be chosen
to be
\begin{equation}
  \hat \epsilon:=\epsilon \text{(3 TeV)}/ M_X, \qquad \hat g_X:=g_X \text{(3
  TeV)}/M_X, \qquad \theta_{sb}. \label{pars}
\end{equation}
One typically wishes to fix the
renormalisation scale $\mu$ of
the Wilson coefficients' boundary condition to be
$M_X$ itself in
order to render unaccounted-for loop corrections small, since they are
expected to be proportional to powers of $\log(M_X/\mu)$. The
renormalisation down to the weak scale $M_Z$ then brings in a separate
dependence upon $M_X$,
although such effects are small, being suppressed by a factor of order
$\log(M_X/M_Z)/(16 \pi^2)$. Since $M_X$ is around the few-TeV scale, it is a
good approximation (at the couple of percent level in relative change to
Wilson
coefficients) to neglect the additional $M_X$
dependence. We shall present results for one particular representative value
of $M_X=3$ TeV but the approximate scaling is implicit because we use the
variables $\hat \epsilon$ and $\hat g_X$ defined in (\ref{pars}).

It was shown by Ref.~\cite{Allanach:2022iod} that the $\hat\epsilon=0$ $B_3-L_2$
model's fit to flavour data became less favourable with LHCb's
reanalysis~\cite{LHCb:2022qnv} of  $R_K$ and
$R_{K^\ast}$ (although it was still more favourable than the SM).
One of the points of our present paper is that kinetic
mixing introduces a
$Z^\prime$-di-electron vertex and can improve the fit.
Therefore, we show the fits for the $B_3-L_2$ mixing with and without including
kinetic mixing (i.e.\ $\hat \epsilon=0$ and $\hat \epsilon \neq 0$,
respectively) in
Table~\ref{tab:best_fits}.
\begin{table}
  \begin{center}
  \begin{tabular}{|ccc|cccc|c|} \hline
    $\hat \epsilon$ & $\hat g_X$ & $\theta_{sb}$ & $\Delta
    \chi^2_\text{quarks}$ & $\Delta \chi^2_\text{EWPO}$ & $\Delta
    \chi^2_\text{LEP2}$ & $\Delta \chi^2_\text{LFU}$ & $\Delta \chi^2_\text{global}$ \\ \hline
    0         & 0.12    &  $-$0.11   &
    35.2      & 0.0     &  $\phantom{+}$0.00    & $-$4.0  & {\bf 31.2} \\
    $-$0.92   & 0.078   & $-$0.17   &
    37.6      & 0.3     & $-$0.02    & $\phantom{+}$0.3   & {\bf \bestchisq} \\
    \hline
  \end{tabular}
  \end{center}
  \caption{Fit to the $B_3-L_2$
    model~\cite{Bonilla:2017lsq,Alonso:2017uky,Allanach:2020kss} excluding
    (top line) and
    including (bottom line) sizeable kinetic mixing. $p$ indicates the
    $p-$value of the best-fit point. $\Delta
    \chi^2:=\chi^2_{\rm SM}-\chi^2$ in each category, so a higher value indicates
    a better fit. Input parameters are shown in the first three columns
    for $M_X=3$ TeV. \label{tab:best_fits}}
  \end{table}
In the top line of the table, we impose $\hat \epsilon=0$ and fit the
remaining two parameters $\{\hat g_X, \theta_{sb}\}$ to show the fit
of the original kinetically unmixed $B_3-L_2$ model.
In the bottom line of the table, we show the
three-parameter $\{\hat \epsilon, \hat g_X, \theta_{sb}\}$ fit to the
malaphoric $B_3-L_2$ model, i.e.\ including
kinetic mixing.
We see from the table that kinetic mixing improves the fit\footnote{The
$p-$value of such a change is $.06$.} by \improvement\ units of
$\chi^2$ at the expense of introducing one additional free parameter
$\hat \epsilon$.
We see
from the table that the improvement over $\hat \epsilon=0$ is mainly due to improving the fit to flavour data
in the `quarks' and to the `LFU' categories of observables.
95$\%$ constraints upon the $\hat \epsilon=0$ $B_3-L_2$ model from flavour
measurements (albeit with previous lepton flavour universality data which
favoured larger new physics effects) were presented in
Refs.~\cite{Allanach:2022iod}.
The malaphoric $B_3-L_2$ model shown in the bottom line has a $\Delta \chi^2$ of
\bestchisq\ in comparison to the SM\@. This is significant: the $p-$value of such
a change, with 3 degrees of freedom (the effective number of free parameters)
is $1. \times 10^{-8}$, indicating that the change is very unlikely to be due
solely to random fluctuations of the measurements.

\begin{figure}
  \begin{center}
    \includegraphics[scale=0.5]{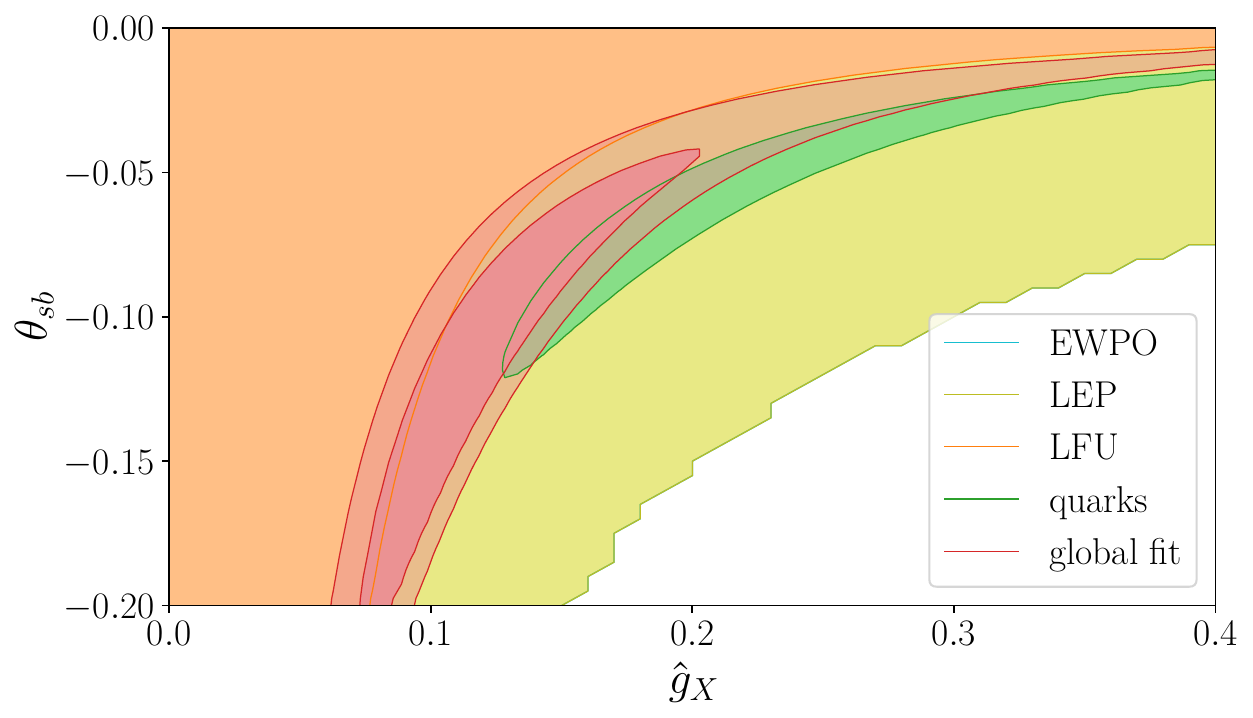}
  \end{center}
  \caption{\label{fig:b3ml2} Parameter space of the unmixed $B_3-L_2$
    $Z^\prime$ model. White regions of parameter space are where
    \flavio\ could not constrain the CKM matrix elements to have an absolute value
    less than or equal to 1; in practice such regions
    are highly excluded by flavour measurements.
    The coloured region shows the 68$\%$ CL allowed
    region for each set of observable (as shown by reference to the legend).
    The global fit region is shown in red and shows both the 68$\%$ CL and the 95$\%$ CL.
  There is no region in the parameter space
    shown where the EWPO data  provide a bad fit and so there is no EWPO contour shown.}
  \end{figure}
In Fig.~\ref{fig:b3ml2}, we show the fit to the $B_3-L_2$ model for
$\hat \epsilon=0$, i.e.\ neglecting kinetic mixing.
We see from the figure that the fit is not perfect because the preferred
region of the `quarks' category of observable does not overlap with the
preferred region of the `LFU' observables at the 68$\%$ confidence level (CL).

\begin{figure}
  \begin{center}
    \unitlength = \textwidth
    \begin{picture}(0.9,0.5)
    \put(-0.05,0){\includegraphics[scale=0.5]{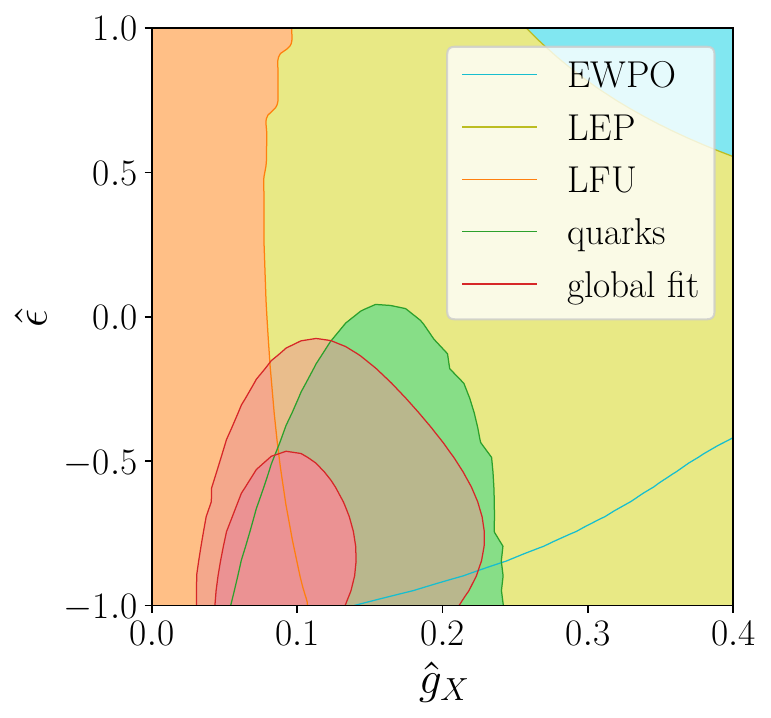}}
    \put(0.45,0){\includegraphics[scale=0.5]{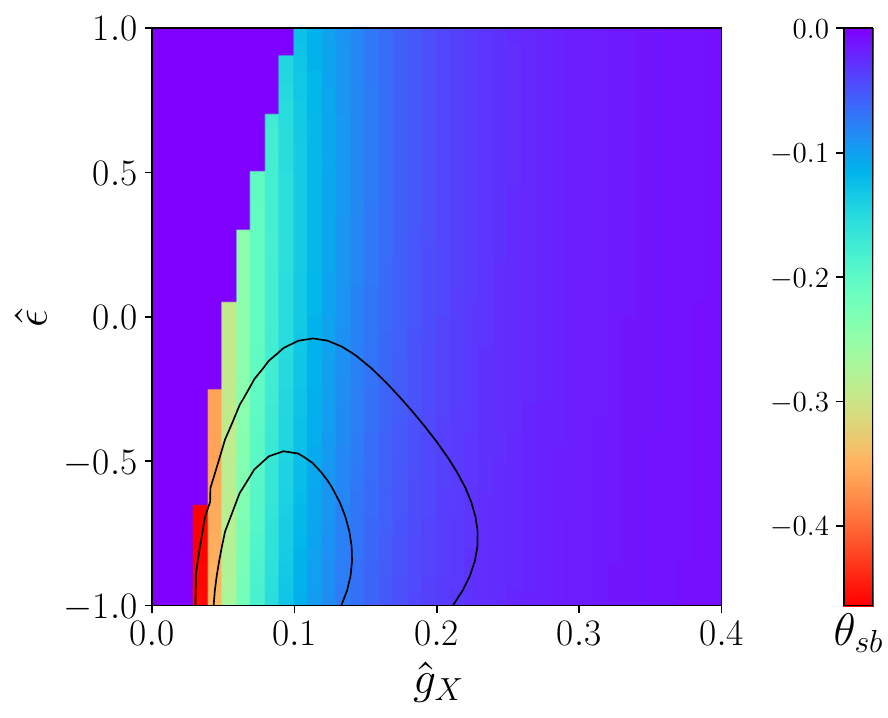}}
    \end{picture}
  \end{center}
  \caption{\label{fig:b3ml2mal} Parameter space of the malaphoric $B_3-L_2$
    $Z^\prime$ model, where $\theta_{sb}$ has been profiled over.
    In the left-hand plot, the coloured region shows the 68$\%$ CL allowed
  region for each set of observable (as shown by reference to the legend).
    The global fit region is shown in red and shows both the 68$\%$ CL and the 95$\%$ CL.
  In the right-hand plot, we show the
  best-fit value of $\theta_{sb}$ across the parameter plane.}
  \end{figure}

We show how this is improved by displaying the parameter space of the \emph{malaphoric} $B_3-L_2$ model in
Fig.~\ref{fig:b3ml2mal}.
We see from the plot that a parameter
region exists which is preferred by the $B-$hadron decay data (`quarks') as
well as LFU observables whilst simultaneously being compatible with LEP2
di-lepton production and measurements of EWPOs.

\begin{figure}
  \begin{center}
    \includegraphics[height=0.4 \textheight]{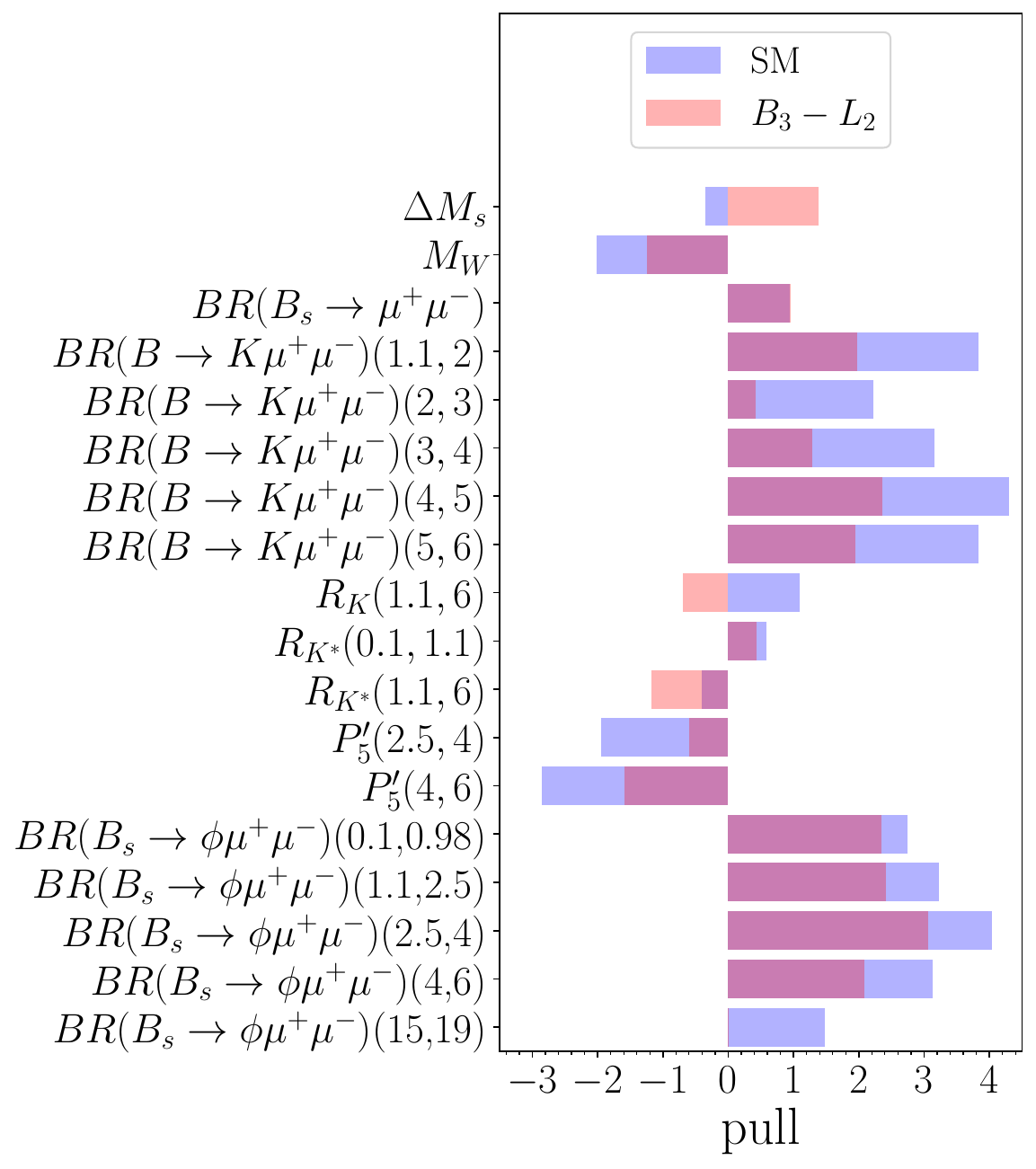}
  \end{center}
  \caption{\label{fig:int} Some observables of interest from \flavio\ for the
    malaphoric
    $B_3-L_2$ model at its best fit point (listed in Table~\ref{tab:best_fits}). The
    `pull' is
    defined as theory prediction minus the
    experimental central value, all divided by uncertainty neglecting any
    correlations. Where an observable has parenthesis with two numerical
    values enclosed, it is for that particular bin in di-muon invariant mass
    squared in units of $\text{GeV}^2$.
}
  \end{figure}

We display some observables of interest for the best-fit point of the
malaphoric $Z^\prime$ model in Fig.~\ref{fig:int}.
We see that many observables are fit better by the malaphoric $B_3-L_2$ model
than the SM,
but a few (notably $\Delta M_s$ and $R_{K^{(\ast)}}(1.1,6)$) have a somewhat worse
fit in the malaphoric $B_3-L_2$ model. $M_W$ at the best-fit point is only
marginally improved with respect to the SM value, which is some 2$\sigma$ off
the measurement. Many of the branching ratios in various bins of di-muon
invariant mass squared are also better fit in the malaphoric $B_3-L_2$
model. We note that in the malaphoric $B_3-L_2$ model, several observables,
although improved upon compared to the SM, are still not fit perfectly. This is reflected
in the $p-$value of the model, which comes out to be\footnote{The $p-$value of
the $\hat\epsilon=0$ $B_3-L_2$ model is $.006$.} $.01$.

We shall now try to gain some approximate qualitative analytic understanding
of some of the
results, neglecting the renormalisation effects (which were taken into
account in our numerical results).
It is known~\cite{Alguero:2023jeh,Hurth:2023jwr} that
in weak effective
theory (WET) language, the fit improvements to
\bsll\ data come dominantly from assuming BSM contributions to $C_{9(\mu)}$,
$C_{10(\mu)}$, $C_{9(e)}$ and $C_{10(e)}$, where we write\footnote{Here, we
have followed the convention of Ref.~\cite{Aebischer:2015fzz}.} the dimension-6 WET
interaction Lagrangian as
\begin{equation}
  {\mathcal L}_{WET} = \frac{4 G_F}{\sqrt{2}}\sum_i (C_i^{\rm SM} + C_i) {\mathcal
    O}_i +
  H.c.,
\end{equation}
where $G_F$ is the Fermi constant, $C_i$ are dimension-6 BSM Wilson
coefficients and of particular interest are the operators
\begin{eqnarray}
  {\mathcal O}_{9(\mu)} &:=& \frac{e^2}{16 \pi^2} \left( \bar s \gamma_\alpha P_L b
  \right) \left( \bar \mu \gamma^\alpha \mu \right), \nonumber \\
  {\mathcal O}_{10(\mu)} &:=& \frac{e^2}{16 \pi^2} \left( \bar s \gamma_\alpha P_L b
  \right) \left( \bar \mu \gamma^\alpha\gamma_5 \mu \right), \label{ops}
\end{eqnarray}
where $e$ is the electromagnetic gauge coupling, $P_L$ is the spinorial left-handed projection operator $(1-\gamma_5)/2$,
$\mu$ is the muon field, $b$ the bottom quark field and $s$ the strange quark
field. ${\mathcal O}_{9(e)}$ and ${\mathcal O}_{10(e)}$ are obtained by
replacing the muon field $\mu$ by the electron field $e$ everywhere in
(\ref{ops}). The same four operators have a version with a prime, where the
projection operator is switched from
$P_L \rightarrow P_R:=(1+\gamma_5)/2$.
In the malaphoric $B_3-L_2$ model presented here, the
matching~\cite{Aebischer:2015fzz} between the SMEFT and the WET
implies that the aforementioned primed
operators all vanish and moreover that
\begin{eqnarray}
  C_{9(\mu)} / K &=&  -\frac{g_X \sin 2 \theta_{sb} }{M_X^2} \left[3  g_X +
    g^\prime  \epsilon \left( s_W^2 - 1\right) \right],
  \nonumber\\
  C_{9(e)} / K &=&  -\frac{g_X \sin 2 \theta_{sb} }{M_X^2}
  g^\prime  \epsilon \left( s_W^2 - 1\right),
  \nonumber\\
    C_{10(\mu)} &=& C_{10(e)} = 0, \label{WET}
\end{eqnarray}
where $s_W:=\sin \theta_W$, $\theta_W$ is the Weinberg angle and
$K:=2\sqrt{2} \pi^2/(e^2 G_F)$; see Appendix~\ref{app:match} for more details.
We see from (\ref{WET}) that, to obtain LFU in \bsll\ transitions i.e.\
$C_{9(\mu)}=C_{9(e)}$, one would require
$
  g_X = 0
$, removing the BSM effects.
However, Ref.~\cite{Allanach:2023uxz}
showed that having $C_{9(e)}=C_{9(\mu)}/2$
is a close-to-optimal fit (to data similar to those taken here) along a particular line
of models. Ref.~\cite{Allanach:2023uxz} also showed, though, that
having $C_{9(e)}=C_{9(\mu)}$ or even $C_{9(e)}=0$ is within the
95$\%$ CL, providing statistical wiggle room.

The branching ratio of $B_s \rightarrow \mu^+ \mu^-$ can receive a BSM
contribution to its SM prediction $BR(B_s \rightarrow \mu^+ \mu^-)^{\rm SM}$ in
$Z^\prime$ models with zero primed operators~\cite{Allanach:2022iod}:
\begin{equation}
  BR(B_s \rightarrow \mu^+\mu^-) = \left| 1 +
  \frac{C_{10(\mu)}}{C_{10}^{\rm SM}}\right|^2 BR(B_s \rightarrow \mu^+\mu^-)^{\rm SM}.
  \end{equation}
The fact that $C_{10(\mu)}=0$ in the malaphoric $B_3-L_2$ model
despite mixing with the hypercharge gauge boson which has chiral interactions\footnote{This can be seen in more detail as
  resulting from a cancellation between the terms proportional to $\epsilon$
  in (\ref{gen}) and (\ref{wetSMEFT}) from
$(C_{lq}^{(1)})^{2223}$, $C_{qe}^{2322}$ and $(C_{\varphi q}^{(1)})^{23}$.},
constrains the $B_s \rightarrow \mu^+\mu^-$ branching ratio to be identical
to the SM prediction, within our approximation. This is borne out at the
best-fit point, as Fig.~\ref{fig:int} shows.

The $B_s-\bar{B}_s$ mixing observable $\Delta M_s$, receives a correction
in the malaphoric $B_3-L_2$ model identical to that of the $\hat \epsilon=0$
limit, namely~\cite{Allanach:2022iod}
\begin{equation}
  \frac{\Delta M_s}{\Delta M_s^{\rm SM}} =
  \left| 1 + \frac{\eta(M_X) {g_X}^2 \sin^2 2 \theta_{sb}}
         {8 R_{\rm SM}^\text{loop} M_X^2}
         \frac{\sqrt{2}}{4 G_F (V_{tb} V_{ts}^\ast)^2}
           \right|, \label{bsbs}
\end{equation}
where
$R_{\rm SM}^\text{loop}=1.3397 \times 10^{-3}$ and $\eta(M_X)$ parameterise
renormalisation effects: $\eta(M_X)$ varies between 0.79 and 0.74 when $M_X$
ranges between 1 and 10 TeV~\cite{DiLuzio:2017fdq}. The SM prediction of
$\Delta M_s$ is in good agreement with the measurement, whereas (\ref{bsbs})
shows
that one acquires a
positive contribution from the malaphoric $B_3-L_2$ model, as shown in
Fig~\ref{fig:int}.
Such a contribution
then has a preference for smaller values of $|\hat g_X \sin 2 \theta_{sb}|$
in the fit.

The prediction of $M_W$ receives a positive contribution from the changes to
$C_{\varphi \Box}$ and $C_{\varphi D}$ in (\ref{Cboxphi}) and
(\ref{CphiD}). As Fig.~\ref{fig:int} shows, a positive contribution is
preferred by the current
measurements but in practice, for the best-fit point at least, such a change
is rather small.

\section{Summary \label{sec:conc}}

We have calculated the dimension-6 SMEFT Wilson coefficients resulting from
integrating out a
kinetically-mixed TeV-scale $Z^\prime$ field
resulting from a spontaneously broken $U(1)_X$
gauge symmetry.
Such Wilson coefficients
were recently
calculated\footnote{Ref.~\cite{Dawson:2024ozw} also
calculates the dimension-8 coefficients, finding that
dimension-8 coefficients are generically too small to affect the phenomenology
much. We neglect them here.} for a generic heavy $Z^\prime$ model including kinetic mixing in
Ref.~\cite{Dawson:2024ozw}. After we found a few sign errors in the
dimension-6
coefficients of the original \emph{pre-erratum}
version of that pioneering paper, the authors of Ref.~\cite{Dawson:2024ozw}
have graciously confirmed that they agree with our expressions, which are
tabulated in (\ref{Cll})-(\ref{CphiD}).

We apply the SMEFT Wilson coefficients mentioned above to the case of a
$Z^\prime$ model (the $B_3-L_2$ model) that was originally designed to fit $b
\rightarrow s \ell^+ \ell^-$ data, which are at odds with their SM predictions.
To our knowledge, this is the first time that kinetic mixing has been taken
into account in
fits to $b\rightarrow s \ell^+ \ell^-$ data.
When kinetic mixing
is small, this $B_3-L_2$
model contains rather too much lepton flavour universality violation and did
not fit more recent analyses of $R_K$ and $R_{K^\ast}$ by the LHCb
Collaboration well. We have shown that including a sizeable kinetic mixing
between the new $U(1)$ gauge boson and the hypercharge gauge boson provides an
improved fit to $b \rightarrow s \ell^+ \ell^-$ measurements. Such a fit comes
about partly because the kinetic mixing contribution respects lepton flavour
universality.
Various papers~\cite{Greljo:2022jac,Ciuchini:2022wbq,Alguero:2023jeh,Hurth:2023jwr}
have pointed out that a LFU contribution to the WET Wilson coefficient $C_9$ as well as an
enhanced contribution to $C_{9(\mu)}$ provides an improved fit. We
see from the pieces proportional to $\epsilon$ in (\ref{WET})
how the kinetic mixing in the malaphoric $B_3-L_2$ model
provides such a LFU contribution.
In addition, we see in (\ref{WET}) that, despite the
fact that the $Z^\prime$ contains a component of the hypercharge gauge boson
(which couples chirally), the resulting propagating eigenstate couples to
left-handed leptons with the same strength that it couples to right-handed
leptons, to a good approximation.

Other approaches to explain neutral current $b \rightarrow s \ell^+ \ell^-$
anomalies after the updated $R_K$ and $R_{K^\ast}$ analyses have been
taken. For example, in Ref.~\cite{Allanach:2023uxz} a model which had $U(1)_X$ charge
    assignments corresponding to
$      X := 3B_3 -L_e -2 L_\mu,   $
    where $L_e$ is first-family lepton
    number, was found to fit the
    flavour measurements including the current ones for $R_K$ and $R_{K^\ast}$
    reasonably well as a whole even neglecting kinetic mixing. The assignment
    $X:=3B_3-L$ where $L$ is lepton
    number, has also been studied~\cite{Greljo:2022jac,Allanach:2023uxz}.
Some authors have introduced several new $U(1)$ gauge bosons
with family-dependent hypercharge-like
assignments~\cite{Davighi:2023evx,FernandezNavarro:2024hnv};
these can have
various
advantages from the
point of view of motivating observed hierarchies in fermion masses.
As remarked in Ref.~\cite{Allanach:2024nsa}, a new gauged $U(1)_X$ model with large kinetic mixing is physically equivalent to another model with no significant
  kinetic
  mixing where a rational multiple of
  hypercharge has been added to each $X$ charge. This could provide a different avenue for ultra-violet model building.

Returning to the unmixed $B_3-L_2$ model, with the assumptions about fermion
mixing detailed in \S~\ref{sec:model}, the $Z^\prime$ couples only very weakly to first
generation quarks. Direct search bounds coming from non-observation of a
di-lepton bump at the LHC collider are then rather weak
because the production
cross-section is doubly suppressed by bottom quark parton distribution
functions, since $b \bar b \rightarrow Z^\prime$ is the dominant
production mode~\cite{Allanach:2021gmj}. When kinetic mixing is introduced,
this may no longer be the
case and it is likely that the direct search bounds will become stronger\footnote{A recent preprint~\cite{Allanach:2024nsa}
confirms this, finding that $M_{Z^\prime}>2.8$ TeV in the 95$\%$ CL global-fit
window.}
because
the malaphoric $B_3-L_2$ predicts a family universal
component to the couplings to up and down
quarks. However, since direct searches are beyond the scope of the present
paper, we shall burn that bridge when we come to it.

\section*{Acknowledgements}
This work was partially supported by STFC HEP Consolidated grants
ST/T000694/1 and ST/X000664/1.
We thank the Cambridge Pheno Working Group and P. Stangl for
discussions and S.~Dawson, M.\ Forslund and M.\ Schnubel for helpful
comparisons of the dimension-6 SMEFT coefficients.
We thank E. Loisa for comments on the manuscript.
BCA thanks CERN for hospitality extended while part of this work was
undertaken.

\appendix
\section{WET Wilson Coefficients in the $B_3-L_2$
  model \label{app:match}}
The tree-level SMEFT to WET coefficient matching formulae are
known~\cite{Aebischer:2015fzz} (we take $C_{ledq}^{2223}$ and
$C_{ledq}^{2232}$ to vanish, as results
from integrating out a $Z^\prime$):
\begin{eqnarray}
  C_{9(\mu)}/K &=& C_{qe}^{2322} + (C_{lq}^{(1)})^{2223} + (C_{lq}^{(3)})^{2223} -
  (1-4s_W^2) [ (C_{\varphi q}^{(1)})^{23} + (C_{\varphi q}^{(3)})^{23} ], \nonumber \\
  C_{10(\mu)}/K &=& C_{qe}^{2322} - (C_{lq}^{(1)})^{2223} - (C_{lq}^{(3)})^{2223} +
  (C_{\varphi q}^{(1)})^{23} + (C_{\varphi q}^{(3)})^{23}, \nonumber \\
  C_{9(e)}/K &=& C_{qe}^{2311} + (C_{lq}^{(1)})^{1123} + (C_{lq}^{(3)})^{1123} -
  (1-4s_W^2) [ (C_{\varphi q}^{(1)})^{23} + (C_{\varphi q}^{(3)})^{23} ], \nonumber \\
  C_{10(e)}/K &=& C_{qe}^{2311} - (C_{lq}^{(1)})^{1123} - (C_{lq}^{(3)})^{1123} +
  (C_{\varphi q}^{(1)})^{23} + (C_{\varphi q}^{(3)})^{23}, \nonumber \\
  C_{9(\mu)}^\prime/K &=& C_{ed}^{2223} + C_{ld}^{2223} -
  (1-4s_W^2) C_{\varphi d}^{23}, \nonumber \\
  C_{10(\mu)}^\prime/K &=& C_{ed}^{2223} - C_{ld}^{2223} + C_{\varphi d}^{23}, \nonumber \\
  C_{9(e)}^\prime/K &=& C_{ed}^{1123} + C_{ld}^{1123} - (1-4s_W^2) C_{\varphi d}^{23}, \nonumber \\
  C_{10(e)}^\prime/K &=& C_{ed}^{1123} - C_{ld}^{1123} + C_{\varphi d}^{23} \label{gen}.
\end{eqnarray}
Since we have set $V_{d_R}$ to be the
3 by 3 identity matrix, $(C_{\varphi d})_{23}=0$ in the $B_3-L_2$ model. Thus,
note that all of the SMEFT coefficients on the right-hand sides of
$C_{9/10(e/\mu)}^\prime$ are zero at tree-level.
From (\ref{Clq1}) and (\ref{Cqe}) and the charges in Table~\ref{tab:b3l2}, we
have
\begin{eqnarray}
  (C_{lq}^{(1)})^{2223} &=& -\frac{1}{M_X^2} \left( -3 g_X + g^\prime
  \epsilon/2 \right)
  \Xi_{23}, \nonumber \\
  (C_{lq}^{(1)})^{1123} &=& -\frac{1}{2M_X^2}  g^\prime \epsilon \Xi_{23},
  \nonumber \\
  C_{qe}^{2322}&=& -\frac{1}{M_X^2} \left(-3g_X + g^\prime \epsilon \right) \Xi_{23},
  \nonumber \\
  C_{qe}^{2311}&=& -\frac{1}{M_X^2} g^\prime \epsilon \Xi_{23} \label{wetSMEFT}
\end{eqnarray}
The only other relevant
non-zero dimension-6 SMEFT coefficient
given by the malaphoric $B_3-L_2$ model
is
\begin{equation}
  (C_{\varphi q}^{(1)})^{23}=\frac{1}{2M_X^2} g^\prime \epsilon \Xi_{23}, \label{Cphiql23}
\end{equation}
as can be seen from
(\ref{Cphiq_1}).
Substituting (\ref{wetSMEFT}) and (\ref{Cphiql23}) into (\ref{gen}), we
arrive at (\ref{WET}).

\section{Integrating out the $X_\mu$ field \label{app:intout}}

To integrate out the $X_\mu$ field we use the path integral formalism.
The part of the generating functional relevant for our discussion reads
\begin{align}
    \label{eq:ZJ}
    Z[J]
    =
    \int [dX_\mu]
    \exp \!\left[
        i\int d^4x \,
        \left(
          -\frac{1}{4} X_{\mu\nu} X^{\mu \nu} + \frac{1}{2} M_X^2
          X_\mu X^\mu - \frac{\epsilon}{2} B_{\mu \nu}X^{\mu \nu}- X_\mu J^\mu
        \right)
    \right],
\end{align}
with the notation introduced in the \S\ref{sec:coeffs} and where $[dX_\mu]$ denotes the integration over all the paths with the correct boundary conditions.
The penultimate term in the equation above can be rewritten as
\begin{align}
    X_{\mu\nu} B^{\mu\nu}
    \equiv
    (\partial_\mu X_\nu - \partial_\nu X_\mu) B^{\mu\nu}
    = - 2 (\partial_\nu X_\mu) B^{\mu\nu}
    \implies 2 X_\mu (\partial_\nu B^{\mu\nu})
    \,,
\end{align}
where the last equality is derived by integration by parts of the action.
We cast (\ref{eq:ZJ}) in the form
\begin{align*}
    Z[J]
    & =
    \int [dX_\mu]
    \exp \!\left[
        \frac{i}{2}\int d^4x\, d^4y\,
        X_\mu(x) \K^{\mu\nu}(x,y) X_\nu(y)
        - i\int d^4x\, X_\mu \hat{J}^\mu
    \right]
    \\ & =
    \int [dX_\mu]
    \exp \!\bigg[
        \frac{i}{2}\int d^4x\, d^4y\,
    \nonumber\\ & \qquad
        \Big(
            X_\mu(x) +
            \int d^4u\,\hat{J}^\alpha(u)  \Delta_{\alpha\mu}(u,x)
        \Big)
        \K^{\mu\nu}(x,y)
        \Big(
            X_\nu(y) +
            \int d^4v\, \Delta_{\nu\beta}(y,v) \hat{J}^\beta(v)
        \Big)
    \nonumber\\ & \qquad
        - \frac{i}{2}\int d^4x\,d^4y\, \hat{J}^\mu(x)  \Delta_{\mu\nu}(x,y) \hat{J}^\nu(y)
    \bigg]\,,
\end{align*}
where
\begin{equation}
\begin{aligned}
    \K^{\mu\nu}(x,y)
    & =
    \delta^{(4)}(x-y)
    \left(
        g^{\mu\nu}(\partial^2 + M_{X}^2) - \partial^\mu \partial^\nu
    \right)
    \,,\\
    \hat{J}^\mu
    & =
    J^\mu + \epsilon (\partial_\nu B^{\mu\nu})
    \,,\\
    \int d^4y \,\K_{\mu\nu}(x,y)\Delta^{\nu\lambda}(y,z) &=
    g_\mu{}^\lambda \delta^{(4)}(x-z)
    \,,\\
    \Delta_{\mu\nu}(x,y)
    &=
    \int
    \frac{d^4k}{(2\pi)^4} \Delta_{\mu\nu}(k)e^{-k(x-y)}
    \,,\\
    \Delta_{\mu\nu}(k)
    &=
    - \frac{1}{k^2- M_{X}^2}
    \left(
        g_{\mu\nu} - \frac{k_\mu k_\nu}{M_{X}^2}
    \right)
    \,.
\end{aligned}
\end{equation}
Performing the replacement $X\mapsto X - \Delta J$, the functional integral over $dX$ can be easily performed (it gives an infinite constant that can be ignored):
\begin{equation}
\begin{aligned}
    Z[J]
    & =
    \exp \!\bigg[
        - \frac{i}{2}\int d^4x\,d^4y\, \hat{J}^\mu(x)  \Delta_{\mu\nu}(x,y) \hat{J}^\nu(y)
    \bigg]
    \\
    & \times
    \int [dX_\mu]
    \exp \!\left[
        \frac{i}{2}\int d^4x\, d^4y\,
        X_\mu(x) \K^{\mu\nu}(x,y) X_\nu(y)
    \right]
    \\
    & \propto
    \exp \!\bigg[
        - \frac{i}{2}\int d^4x\,d^4y\, \hat{J}^\mu(x)  \Delta_{\mu\nu}(x,y) \hat{J}^\nu(y)
    \bigg]\,.
 \end{aligned}
\end{equation}
Finally, expanding in the limit $M_{X}^2\gg k^2$, we obtain
\begin{align}
    \Delta_{\mu\nu}(x,y)
    &\simeq
    \frac{g_{\mu\nu}}{M_{X}^2}
    \delta^{(4)}(x-y)
    \,,
\end{align}
and hence
\begin{align}
    \label{eq:LagDaw}
    \mathcal{L}_6
    =
    - \frac{1}{2M_{X}^2} \hat{J}^\mu  \hat{J}_\nu
    \equiv
    -\frac{1}{2 M_{X}^2} {J}_\mu {J}^\mu -
    \frac{\epsilon}{M_{X}^2} (\partial_\nu B^{\mu \nu}) {J}_\mu -
    \frac{\epsilon^2}{2 M_{X}^2} (\partial_\nu B^{\mu \nu}) (\partial^\rho
    B_{\mu \rho}),
\end{align}
which coincides with the first line in Eq.~(9) of Ref.~\cite{Dawson:2024ozw}.
The same result can be achieved using the equations of motion for the $X_\mu$
field:
\begin{align}
    &\partial_\nu X^{\mu\nu} = - J^\mu - \epsilon \partial_\nu B^{\mu\nu} + M_{X}^2 X^{\mu}
    \,,\\
    \implies &
    X^\mu = \frac{1}{M_{X}^2}
    \left(
        \partial_\nu X^{\mu\nu} + J^\mu + \epsilon \partial_\nu B^{\mu\nu}
    \right)
    \,.
\end{align}
Substituting this expression into (\ref{eq:LXB}) gives (\ref{eq:L6}).

\bibliographystyle{JHEP-2}
\bibliography{kin}

\end{document}